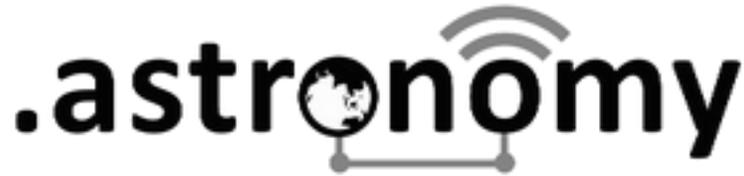

.Astronomy 4 Unproceedings
Heidelberg, July 2012

*Authors: Robert Simpson (Oxford University), Chris Lintott (Oxford University), Amanda Bauer (Australian Astronomical Observatory), Bruce Berriman (IPAC, Caltech), Edward Gomez (LCOGT), Sarah Kendrew (MPIA), Thomas Kitching (MSSL/UCL), August Muench (Harvard CfA), Demitri Muna (New York University), Thomas Robitaille (MPIA), Megan E. Schwamb (Yale University), Brooke Simmons (Oxford University)*

**Abstract**

The goal of the .Astronomy conference series is to bring together astronomers, educators, developers and others interested in using the Internet as a medium for astronomy. Attendance at the event is limited to approximately 50 participants, and days are split into mornings of scheduled talks, followed by 'unconference' afternoons, where sessions are defined by participants during the course of the event. Participants in unconference sessions are discouraged from formal presentations, with discussion, workshop-style formats or informal practical tutorials encouraged. The conference also designates one day as a 'hack day', in which attendees collaborate in groups on day-long projects for presentation the following morning. These hacks are often a way of concentrating effort, learning new skills, and exploring ideas in a practical fashion. The emphasis on informal, focused interaction makes recording proceedings more difficult than for a normal meeting. While the first .Astronomy conference is preserved formally in a book, more recent iterations are not documented. We therefore, in the spirit of .Astronomy, report 'unproceedings' from .Astronomy 4, which was held in Heidelberg in July 2012.

---

1. Introduction to .Astronomy

The .Astronomy series of conferences began in Cardiff in 2008, followed by Leiden in 2009, Oxford in 2011 and, most recently, Heidelberg in 2012. The goal is to bring together astronomers, educators, developers and others interested in using the Internet as a medium for astronomy, whether for research, for outreach or otherwise. Attendance is limited to approximately 50 participants, a number large enough to inspire productive group work, but not so large that participants can hide in anonymity. Everyone can and does contribute. The conferences usually follow the same, perhaps unusual, format: on the first and third days, scheduled talks in the morning are followed by an 'unconference' of sessions defined by the participants during the course of the conference. Participants in the unconference sessions are (strongly, and sometimes forcibly) discouraged from formal presentations, with discussion, workshop-style formats or informal practical tutorials encouraged. The second day of the three-day conference is mostly taken up by 'hack day', in which attendees collaborate on day-long projects for presentation the next day. These hacks are often a way of concentrating effort, learning new skills, and exploring ideas in a practical fashion.

The emphasis on informal, focused interaction makes for a lively conference, but makes recording proceedings more difficult than for a normal meeting; while the first .Astronomy conference is preserved formally in a book (ISBN 0–9549846–9–2) more recent iterations which used the current format are not documented. We therefore, in the spirit of .Astronomy, report 'unproceedings': a document edited by attendees that aims to give some sense of the state of astronomy on the web in 2012; highlighting areas of particular interest or those where further attention is needed.

Of the 51 attendees at .Astronomy 4, 37 were representing universities and several were not professional astronomers in their working lives. 16 were women and 35 were men; 18 were based in the Unites States, 2 in South Africa, 1 in Australia and the rest in Europe. All attendees share an interest in technology; for example, the ratio of web-enabled devices to people is usually in excess of 2 - with 48 of the 51 attendees in 2012 having Twitter accounts, even if some are more active than others. The use of Twitter is pervasive during the conference; more than 1200 tweets used the #dotastro hashtag during the three-day meeting, with more than 200 tweeters participating in discussion - showing that Twitter may be an excellent way to expand the .Astronomy audience in realtime. These tweets are preserved (at http://eventifier.co/event/dotastro/tweets) but are somewhat uneven; for example, an unconference session about Twitter was notably absent on Twitter itself, due to the distraction of the participants from their favourite medium. A summary recorded (in blog form) by Karen Masters (University of Portsmouth, http://twitter.comKarenLMasters) and can be found online at http://thebeautifulstars.blogspot.co.uk/2012/07/twitter-recommendations-from.html.

.Astronomy 4 was hosted by the Max Planck Institute for Astronomy (MPIA) at the Haus der Astronomie and the Internationales Wissenschaftsforum Heidelberg (IWH). The local organizing committee were Sarah Kendrew (MPIA, https://twitter.com/sarahkendrew), Tom Robitaille (MPIA, https://twitter.com/astrofrog) and Markus Pössel (MPIA, https://twitter.com/mpoessel), and the Scientific organization committee was led, as ever, by Robert Simpson (Oxford University, https://twitter.com/orbitingfrog).

Details of the event's participants, talks and hack day projects can be found online at http://dotastronomy.com/events/four/.

---

## 2. Visualization

A deliberate focus of the 2012 .Astronomy was on visualization. In particular, Noah Illinsky (Author, https://twitter.com/noahi) & Julie Steele (O'Reilly Media, https://twitter.com/

[jsteeleeditor](jsteeleeditor)) (authors of 'Beautiful Visualization', ISBN 1-4493-7986-9) presented a keynote which emphasised the need for careful choice of tactics and strategy in developing memorable images, both for presentations and in journal publications.  Their talk left many with a uneasy feeling that we should revise every plot we've ever made. Underlining the essential need to consider data, audience and design, the message was to focus on essentials rather than adding extraneous information at the expense of clear communication. Some guidelines, such as the fact that human perception is bad at distinguishing more than about twelve gradations in angle - making pie charts particularly ineffective for complex data - are well known and still often ignored, whereas others were new to us.

A "golden rule" was presented:  'location is everything, colour is difficult' - which is overlooked by almost everyone, resulting in plots which communicate much to their proud creators, and little to everyone else. One example of a good visualization highlighted in the presentation - the flight display at [http://www.hipmunk.com](http://www.hipmunk.com), which rapidly displays and compares a large number of flight options - will be of interest to astronomers for more than just its excellent information design.

One solution is to ask others for help while developing visualisations, and the availability of a wide range of tools both to make visualizations and to give advice on colour choice make the task easier. A series of unconference sessions which operated as 'surgeries' for those with sickly visualizations were very effective, allowing the group to critique and improve each other's work, and would be an excellent addition to any conference, and could also be implemented by groups or departments.

*Summary*

- *Visualisation of data in both papers and presentations is a frequently overlooked aspect of communicating scientific results.*
- *Visualisation consultants, surgeries and sessions could be extremely valuable for improving astronomers' skills sets in this area.*

*Relevant video links:*
Dataverse ([http://dotastronomy.com/events/four/astronomy-dataverse-august-muench/](http://dotastronomy.com/events/four/astronomy-dataverse-august-muench/))
Mechanical Turk, Visualisation and 'Glue' ([http://dotastronomy.com/events/four/mechanical-turk-and-glue-michelle-borkin/](http://dotastronomy.com/events/four/mechanical-turk-and-glue-michelle-borkin/))
Visualisation ([http://dotastronomy.com/events/four/how-to-be-a-data-visualization-star-julie-steele-and-noah-iliinsky/](http://dotastronomy.com/events/four/how-to-be-a-data-visualization-star-julie-steele-and-noah-iliinsky/))

---

### 3. Javascript, Python and astronomy in the browser

A novel feature of the 2012 .Astronomy was the drive amongst some attendees to build serious tools for data manipulation in the browser. Amit Kapadia (Adler Planetarium, [https://twitter.com/a_kapadia](https://twitter.com/a_kapadia)) presented astro.js ([http://www.astrojs.org/](http://www.astrojs.org/)), an ongoing project to create a library of useful tools. The motivation for this change is twofold. Firstly, the advent of widespread HTML5 in browsers and the development of Javascript technologies that improve the ease with which code can be written by those used to other languages make such an approach much more tractable than it would have been in the past. Coffeescript, for example, makes the experience of writing Javascript similar to that of coding in Python or Ruby, while Javascript frameworks such as SpineJS take care of much of the heavy lifting.

Secondly, and slightly more controversially, many attendees felt that data manipulation in the browser provided a place for `play', i.e. for the rapid exploration of new datasets and images. The aim is not to *replace* but *reduce* the need for specialist desktop tools, so that initial exploration can be carried out without ever downloading data or getting to grips with software. A third motivation might be that browser-based tools, particularly now that reliance on Flash is much less prevalent than even a couple of years ago, are inherently accessible to a wider community. If this is the case, we should expect to see more educators, citizen scientists, amateur astronomers and others using Javascript tools in the future.

As a demonstration, Kapadia showed fits.js, a tool which allowed the manipulation of a FITS image in the browser. The entire image is passed to the browser, where it can be rescaled, stretched or otherwise altered. The advantage of this approach is primarily in the reduction of the need for compute capacity on the server side, producing a system which potentially scales more easily and, critically, more economically. It is also possible to conduct calculations rapidly; the 'Lens Toy' demonstrated by Stuart Lowe (LCOGT, https://twitter.com/astronomyblog) predicts strong lensing by a cluster with four components which updates as fast as the mouse can be moved. This tool, which could be extended further to provide rapid modeling of lens candidates, exemplifies the emerging strategy of building tools which can be used for rapid assessment of interesting finds. It is interesting to note that this last tool actually started life as a hack at .Astronomy 3 in Oxford, 2011.

During the hack day, this approach was extended beyond image manipulation further via a series of demonstration projects. Daniel Foreman-Mackay (NYU, https://twitter.com/__dfm__) demonstrated optimization in the browser; this optimization was applied to perform a rapid a Gaussian line fit in the browser, thus highlighting the potential for Javascript as a scientific programming language. Given data encoded as a JSON string (use of JSON to pass data was another commonly used technology during hack day), a Gaussian fit of arbitrary order is applied in response to user input - a demo exists at http://dan.iel.fm/optimize.js/examples/gaussfit/. Meanwhile several participants built Flora, a tool which allows for photometry of an image, or of sections of an image highlighted by a user (https://github.com/kapadia/Flora).

These hack day projects demonstrated that Javascript is now a mature technology for astronomical purposes, capable of being used for rapid development of interesting tools. As one participant noted, however, it is not yet clear whether this is yet another language the astronomical developer will be required to learn, as adoption of Javascript may be slower than progress toward a simple means of porting Python into Javascript.

Python was the most popular language for work listed by 42% of attendees who responded to an informal survey (see the summary at http://prezi.com/icuno6aja8in/dotastronomy-survey/

). IDL came second (18%) and Javascript third (13%). An extremely large number of Python libraries aimed at astronomers is now available, many of them with overlapping aims and capabilities. An attempt to produce a large, standard library (Astropy) has reached its 0.1 release, and is available at http://www.astropy.org.

Before we rush to declare either Javascript or Python the sine qua non of future astronomical analysis, it is worth noting the large proportion of .Astronomy attendees who listed IDL as their language of choice (to the surprise and, perhaps, outrage of many a Python developer in attendance[1]). This result shows that licensed, yet powerful, languages still have a place in the heart of many. Given the high cost of IDL licenses - and many other, similar professional software packages - it is perhaps worth some work to understand whether this preference is due simply to people's habits or education, or maybe that they have invested time writing complex code in one language and do not wish to port it to another. Perhaps it is the ease of use afforded by a given language in their working environment, or in rare cases, functionality that cannot be found anywhere else.

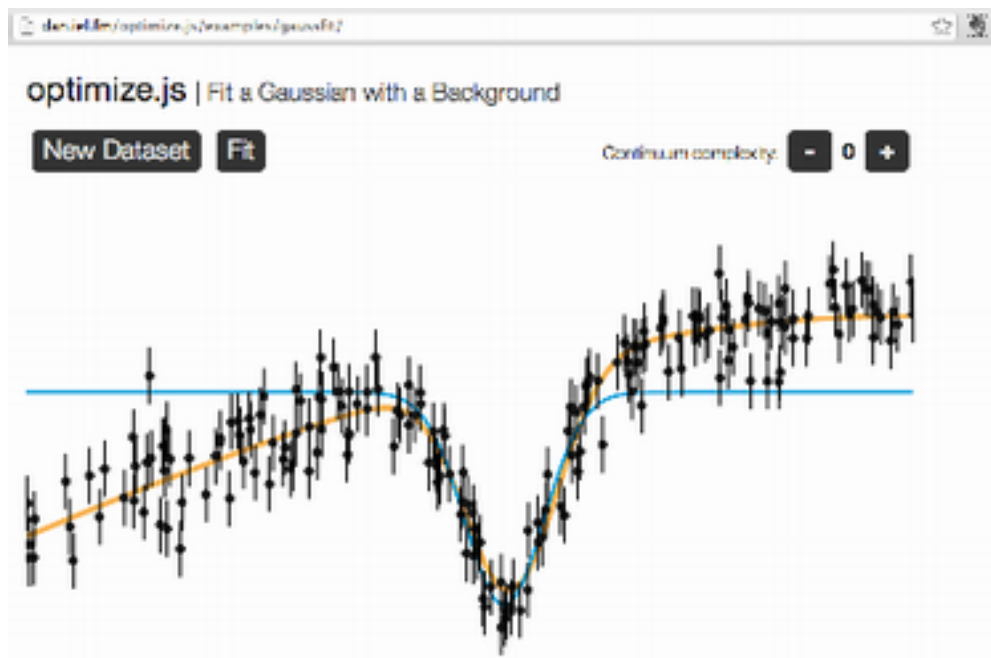

In-browser data fitting using Javascript

## Summary

- *.Astronomy is a good environment to study astronomers' latest coding behaviours.*
- *Python is increasingly becoming the baseline choice for programming. IDL remains very popular.*

---

[1] A diversity of opinion recorded in this outreach video, another hack day activity, http://www.youtube.com/watch?v=M-AzxQeTaw4

- *Javascript is a powerful language for in-browser analysis and visualisation tools, allowing responsive analysis of data stored on remote servers.*

*Relevant video links:*
AstroJS (http://dotastronomy.com/blog/2012/08/amit-kapadia-Javascript-for-astronomy/)
Astropy(http://dotastronomy.com/events/four/astropy-thomas-robitaille/)

---

## 4. Crowdsourcing

Modern astronomy is beset with a data problem. Bruce Berriman (IPAC, https://twitter.com/bruceberriman) noted that we may have, on a very conservative estimate, more than 120TB of data available to us by the end of the decade, and a substantial thread running through the conference was the need to get some help in dealing with this. Crowdsourcing in astronomy has primarily been via the now familiar microtasking of non-experts, but Thomas Kitching (MSSL/UCL, https://twitter.com/tom_kitching) has managed to find true expertise amongst the crowd. Using Kaggle (http://www.kaggle.com), one of a growing number of platforms for machine learning and computer vision challenges (see also Crowdflower and, for NASA-specific projects, NASA Tournament Lab), his collaborators asked for help with algorithms for weak lensing. Inspired by the chance to win a trip to JPL to meet the astronomers (!), rapid progress was made in only two months of competition, which improved on the performance of existing astronomer-built solutions. Solutions which scored highly came from participants with a wide range of backgrounds, including a glaciologist who adapted a routine originally used for identifying icebergs[2]. Encouragingly, the top three solutions each employed very different approaches, suggesting many further avenues for future improvement.

Microtask crowdsourcing is still extremely useful for many astronomical tasks, having being deployed most successfully by the Zooniverse collaboration (https://www.zooniverse.org/). One particularly novel twist was the development during hack day of 'Zoonibot', an artificial participant in discussion on the planethunters.org forums, where many members of the public ask questions about exoplanets and the Kepler mission data shown on the main Planet Hunters site (http://planethunters.org). This auto-posting code was able to respond to requests for help, correct common misunderstandings and participate in discussion (https://github.com/adrn/ZooniBot). While much further work is required to give Zoonibot a modicum of intelligence (an ability to discriminate new from experienced volunteers is particularly desired), it is pleasing to say that it has been accepted by the community of citizen scientists, one of whose children supplied a drawing of Zoonibot's imagined physical form less than an hour after its appearance.

The development of projects such as those in the Zooniverse requires substantial development effort, and a talk by Michelle Borkin (Harvard, https://twitter.com/michelle_borkin) illustrated the use of Amazon's Mechanical Turk as a crowdsourcing platform, in this case investigating human perception of dendograms as part of an investigation of visualization. The system is easy to set up, and extremely rapid as `turkers' complete tasks in exchange for often small monetary rewards. (One conference attendee was very proud of earning four cents during the talk by carrying out tasks on Turk). For projects which would benefit from crowdsourcing but which do not need the scale of effort provided by larger, popular platforms, Turk looks like an affordable and easy option. (A comparison of accuracy between volunteers and turkers working for money would, however, be fascinating!).

---

[2] Incidentally we suggest that this flow of expertise could be reciprocated: astronomers may be able to supplement their grant funding by adapting our routines to help other researchers in their domains.

*Summary*

- *Crowdsourcing can provide effective solutions to certain big data challenges.*
- *Online crowdsourcing platforms can drive cross-pollination between scientific disciplines.*
- *Two-way engagement can be made more effective with well designed artificial intelligence tools.*
- *Small monetary rewards may be beneficial for some crowdsourcing tasks.*

**Relevant video links**

Astronomy Crowdsourcing (http://dotastronomy.com/events/four/astro-crowdsourcing-thomas-kitching/)
Mechanical Turk, Visualisation and 'Glue' (http://dotastronomy.com/events/four/mechanical-turk-and-glue-michelle-borkin/)

---

**5. Career structure and development**

.Astronomy deliberately attempts to recruit participants with a range of technical abilities, partly as a means of increasing the level of such skills and the confidence to deploy them in the community. Hack day in particular acts as a safe and supportive space to encounter new technologies and to learn from each other, and was particularly successful in engaging most of us late into the night with a wide range of projects. It highlights the importance of "play" in research, with many successful hacks originating in "what if...." discussions. We feel that employers should allow and indeed *stimulate* researchers, particularly those in the early stages of their careers, to pursue unconventional ideas and engage with a broad science community beyond their immediate field.

Unconference sessions were devoted to exploring how this strategy might be deployed elsewhere, perhaps at AAS or other large conferences (Indeed, a hack day was organized by .Astronomy alumni at the 221st AAS Winter Meeting in January 2013, and a stand-alone hack day event took place in New York in December 2012). Whether this is successful in the long term or not, we suggest that the behaviours exhibited at hack day, particularly pair- or group-coding, with groups including those with a range of abilities and skills, would be an extremely effective way of increasing the abilities of an entire research group, and would be a useful alternative to the traditional academic practice of working separately on different aspects of the same program.

Information sharing is also important, and AstroBetter (http://www.astrobetter.com), represented by Kelle Cruz (CUNY, http://www.hunter.cuny.edu/physics/faculty/cruz), has established itself as an information hub for astronomers. Attendees (and the wider community) were encouraged to submit articles. Recently, AstroBetter has focused on wider issues beyond the technical, with an emphasis on the importance of mentoring and what seems to be widespread 'impostor syndrome' in astronomy (although the evidence is anecdotal at the time of writing). An unconference session led by Emily Rice (CUNY, http://research.amnh.org/astrophysics/staff/erice) attempted to identify technical solutions to better present such information, focussing on ways of targeting a broad audience.

Bruce Berriman led an unconference session on the best ways of providing software engineering training for astronomers (http://astrocompute.wordpress.com/2012/08/10/software-training-discussion-session-at-astronomy-4/).

*Summary*

- *The .Astronomy Hack Day demonstrates the effectiveness of pair- or group-coding and collaborating in small multi-disciplinary teams.*
- *All researchers should be encouraged to spend time pursuing unconventional ideas and developing their own unique talents.*
- *Social networking platforms such as blogs play an important role in building a global community of researchers, offering pastoral support as well as technical/scientific advice.*

**Relevant video links:**
AstroBetter (http://dotastronomy.com/events/four/astrobetter-kelle-cruz/)

---

### 6. Literature hacking

A recurring theme for the last few .Astronomy conferences has been in novel ways to interface with the astronomical literature, a key feature of the vision of 'seamless astronomy' promoted by Alyssa Goodman (Harvard, https://www.cfa.harvard.edu/~agoodman/). The recent development of ADS Labs (http://adslabs.org) represents a mainstream deployment of discovery tools based on literature mining, but more could be done. One hack day project by Robert Simpson, Karen Masters (Portsmouth, https://twitter.com/KarenLMasters) and Sarah Kendrew looked at the shared language between connected papers, demonstrating the perceived importance of galaxies to the study of dark energy, for example.

Developments in accessing and exploring the scientific literature are not limited to astronomy, and a wide range of recent tools were reviewed in unconference sessions on the topic. Initiatives such as http://www.figshare.com (data sharing), http://www.f1000.com (paper recommendation from experts) are competing with many others (e.g. http://datadryad.org/, http://www.plosone.org/home.action, http://www.myexperiment.org/) for what they clearly expect to be a large audience eager for new ways of doing things. Returning to our own area, the Astronomy Dataverse Network (presented by August Muench (CfA, https://twitter.com/augustmuench), http://www.theastrodata.org) attempts to collate together a research groups' data into a searchable and linkable set. Changes to peer review and publication were, as always, central to these discussions, and Peter Melchior (Ohio State, https://twitter.com/openpaperreview) presented http://www.paperrater.org, a proposed open review platform for astronomical papers. Alberto Pepe (Harvard CfA, http://twitter.com/albertopepe) described the ADS All-Sky survey, an automated extraction of the NASA ADS data to produce explorable sky maps of objects in the literature and go further, showing data from those publications in a linked form.

Regardless of the technology used, there seemed to exist a strong consensus around reclaiming 'publishing' from the peer reviewed paper, in favor of exposing the full output of scientific work, including code, data, workflows and even dead ends.

*Summary*

- *Data isn't just numbers: online literature databases and repositories offer rich opportunities for interacting with and discovering the literature in innovative ways.*
- *Such new literature exploration methods are being explored at many levels: by individual researchers, institutions and technology companies, suggesting both real research impact and the potential for generating revenue.*
- *The web is an ideal platform for exploration of non-traditional peer review strategies.*

---

**7. Education for a global audience**

The .Astronomy attendees were once again overwhelmingly based at European and American institutions (the latter making a particularly strong showing this time), yet astronomy is an inherently global activity (anyone can look up at the sky at night) and technology remains a way of reaching a distributed audience. The SKA bid by South Africa in particular highlighted the transformative effect astronomical infrastructure can have, but care is needed for web projects to global communities. As well as the problem of translations, for much of the world the Internet is primarily accessed through mobile devices. With care, however, this design challenge can be met and an audience beyond astronomy's traditional centers can be engaged. Kevin Govender (IAU, https://twitter.com/govender)  talked about "Astronomy for a better world" in the context of his work with the IAU's Office of Astronomy for Development, in South Africa. Their goal is to use astronomy in schools, universities and with the public, to inspire developing parts of the World. With tools available through the Internet, now more than ever, is this achievable on a large scale.

A distributed community can also be exploited, and a large hack day effort was in the development of 'Astro glocal', a system for connecting amateurs and professionals to potentially anyone around the world who would like help with outreach. Further development is expected!

*Summary*

- *Science and scientific education is a powerful development tool, but initiatives must be tailored to local situations.*
- *Astronomy infrastructure can have a transformative effect on the perception of science in a society.*

*Relevant video links:*
Astronomy for a Better World (http://dotastronomy.com/events/astronomy-4/astronomy-for-a-better-world-kevin-govender/)

### 8. Tools and Hacks

This being an unproceedings, rather than a comprehensive record of the conference, we have concentrated on topics which crossed many sessions and are of broad interest. Other topics covered included cloud computing for astronomy, which Bruce Berriman (IPAC) summarized as being powerful for processing and memory bound apps, and useful for high-burst or rapid scaling, but not necessarily cost-effective for long term use. This led to heated debate between several attendees who found themselves eventually in violent and passionate agreement that some, but not all tasks, were suitable for online cloud-computing environments (see http://astrocompute.wordpress.com/2012/07/10/how-can-we-use-cloud-computing-in-astronomy-astronomy4-keynote/).

Tools produced by various VO efforts were also in evidence, from the US Virtual Astronomical Observatory (VAO) Data Discovery Tool[3] to advocates of SAMP[4] and its ability to connect a wide variety of programs, passing data between them. Knowledge and adoption of these tools remains remarkably low amongst the wider astronomical community.

The excellent facilities of the Haus der Astronomie were exploited by attendees during the 24 'Hack Day'. Hackers keen to produce video projects, spent the day and most of the night producing videos aimed at explaining exactly what astronomers really think.

One of the most immediately useful tools was the conversion of Ned Wright's well-known cosmology calculator (http://www.astro.ucla.edu/~wright/CosmoCalc.html) into a dashboard widget for Mac OSX, accepting long lists of input parameters and exporting data to the clipboard. This was created by Brooke Simmons (Oxford, https://twitter.com/vrooje) and Stuart Lowe and it can be obtained here : http://dotastronomy.com/blog/2012/08/cosmology-calculator-os-x-widget/.

---

[3] http://www.usvao.org/science-tools-services/vao-tools-services-data-discovery-tool/

[4] http://wiki.ivoa.net/twiki/bin/view/IVOA/SampInfo

The Planet Hunters assistant 'Zoonibot', developed by several attendees, is explained in a blog post at http://blog.planethunters.org/2012/08/16/zoonibot/. The excellent 'Flora' and 'optmize.js' Javascript hacks can be played with at http://ubret.s3.amazonaws.com/dotastro4/index.html#/examine/m101 and http://dan.iel.fm/optimize.js/examples/gaussfit/ respectively.

A complete list of hacks, with links, can be found online at http://dotastronomy.com/events/four/ .

*Summary*

- *The .Astronomy Hack day demonstrates convincingly that astronomers with a variety of skills can generate ideas and collaborate successfully on projects that go far beyond their traditional day-to-day work.*

---

**9. Conclusion**

.Astronomy 2012 showed that there is a growing community of astronomers using the web to accelerate the pace and effectiveness of their research, and their outreach. It also showed that technologies and tools are changing rapidly, and a flexible approach is necessary to take advantage of opportunities presented by evolving technology. This is a challenge for those planning for the next generation of facilities, but a fabulous chance for the rest of us to get our hands dirty and experiment. .Astronomy 5 will be held in September 2013 (details at http://dotastronomy.com/events/five/); much of what was deemed cutting edge at .Astronomy 4 may then be commonplace, and what will then seem cutting edge may yet to be started.

Visit http://dotastronomy.com for more information and past and future events, and to read more from the community.

---

**10. Acknowledgements**

The .Astronomy organisers received very generous financial support from a variety of sources, without whom the conference would not have been possible. The committee and local organisers would therefore like to express their warmest thanks to our supporters.

**Max Planck Institute for Astronomy, Heidelberg**
**Haus der Astronomie, Heidelberg**
**Deutsche Forschungsgemeinschaft (DFG)**